\documentclass[12pt, preprint]{aastex}

\usepackage{subfigure,epsfig,url}

\begin{document}

\title{55 Cancri: A Coplanar Planetary System that is Likely Misaligned with its Star}

\author{Nathan A. Kaib\altaffilmark{1,2}, Sean N. Raymond\altaffilmark{3,4} \& Martin J. Duncan\altaffilmark{1}}

\altaffiltext{1}{Department of Physics, Queen's University, Kingston, ON K7L 3N6, Canada, nkaib@astro.queensu.ca}
\altaffiltext{2}{Canadian Institute for Theoretical Astrophysics, University of Toronto, Toronto, ON M5S 3H8, Canada}
\altaffiltext{3}{Universit\'e de Bordeaux, Observatoire Aquitain des Sciences de l'Univers, 2 rue de l'Observatoire, BP 89 33271 Foirac Cedex, France}
\altaffiltext{4}{CNRS, UMR 5804, Laboratoire d'Astrophysique de Bordeaux, 2 rue de l'Observatoire, BP 89, 33271, Floirac Cedex, France}

\begin{abstract}

Although the 55 Cnc system contains multiple, closely packed planets that are presumably in a coplanar configuration, we use numerical simulations to demonstrate that they are likely to be highly inclined to their parent star's spin axis.  Due to perturbations from its distant binary companion, this planetary system precesses like a rigid body about its parent star.  Consequently, the parent star's spin axis and the planetary orbit normal likely diverged long ago.  Because only the projected separation of the binary is known, we study this effect statisitically, assuming an isotropic distribution for wide binary orbits.  We find that the most likely projected spin-orbit angle is $\sim$50$^{\circ}$, with a $\sim$30\% chance of a retrograde configuration.  Transit observations of the innermost planet -- 55 Cnc e -- may be used to verify these findings via the Rossiter-McLaughlin effect.  55 Cancri may thus represent a new class of planetary systems with well-ordered, coplanar orbits that are inclined with respect to the stellar equator.  

\end{abstract}

\section{Introduction}

The 55 Cancri planetary system contains five known planets in low-eccentricity orbits between 0.01 and 5.75 AU \citep{but97,mar02,mcarth04,fisch08}. Recently, the innermost planet in the system -- the $\sim$8 $M_\oplus$ 55 Cnc e -- was found to transit \citep{winn11, dem11}.  This is particularly intriguing because it may allow observers to measure (among other properties) the spin-orbit angle of the planetary system via the Rossiter-McLauglin effect \cite[see][]{gaudwinn07}. 

Rossiter-McLauglin measurements have revealed that the orbits of many hot Jupiters are highly inclined relative to their host star's spin axis and are even retrograde in some cases \cite[e.g., ][]{and10, tri10, winn10}. This conflicts with standard models of planet migration \citep{lin96} because protoplanetary disks are generally thought to be aligned with the stellar equator, as has been confirmed by debris disk measurements \citep{wat11}.  The origin of highly inclined or retrograde planetary orbits may therefore be dynamical.  Several recent papers have shown that such orbits may be produced by tidal circularization of highly eccentric/inclined planets after a phase of planet-planet scattering \citep{nag08} and/or Kozai interactions with a binary star or massive planet \citep{fabtre07, naoz11, katz11, lith11}.  

Another noteworthy feature of 55 Cnc is that it also has a 0.27 M$_{\sun}$ binary companion at a projected distance of 1065 AU \citep{mug06}.  While there have been many dynamical studies of this planetary system \cite[e.g., ][]{ray06, bar07, fisch08, ray08}, none have investigated the effects of the binary companion.  Presumably, this is because this well-ordered system displays none of the excited orbits expected from severe impulses or Kozai interactions with its binary companion.  Here we show that this binary companion does in fact play a major role in the planetary system's dynamics.  Using numerical simulations, we demonstrate that it is very likely this binary has driven the planetary system to a very high (and possibly retrograde) spin-orbit angle with respect to the parent star's spin axis.  Moreover, the Kozai oscillations studied in recent works do not produce this this high spin-orbit angle.  Rather, it is a mechanism first described in \citet{inn97}, where the entire planetary system smoothly precesses as a rigid body.

\section{Numerical Methods}
To model the dynamics of the 55 Cancri system, we use the wide binary algorithm of the MERCURY integration package \citep{cham99,cham02}.  All of our simulations include both the primary and secondary stars of 55 Cnc as well as one or more of the system's known planets.  In each simulation, the primary's mass (55 Cnc A) is set to 0.905 M$_{\sun}$ \citep{vonb11}, while the secondary's mass (55 Cnc B) is fixed at 0.27 M$_{\sun}$ \citep{mug06}.  In total, we perform 501 simulations.  450 of our runs include 55 Cnc d, the most distant and massive planet in the system, as well as a hypothetical Saturn-mass planet we designate as 55 Cnc g.  Our motivation for using such systems is discussed in Section 3.  These simulations are integrated for 10.2 Gyrs, as this matches the most recent age estimate of the system \citep{vonb11}.  These runs are divided into nine 50-simulation subsets.  In each subset, 55 Cnc B is assigned a different semimajor axis: 750, 1000, 1250, 1500, 2000, 3000, 4000, 5000 or 8000 AU.  All of its other orbital elements are drawn randomly from an isotropic distribution.  

In our remaining simulations, we repeat some of our 2-planet simulations using a different planetary configuration.  For 48 of these reruns, we use the four outermost planets (55 Cnc b, c, f, and d) in place of 55 Cnc g and d.  These simulations are only run for 1 Gyr because the  orbit of 55 Cnc b necessitates a timestep of just 1 day.  Finally, we also rerun 3 of these simulations using all 5 planets.  These 5-planet simulations are run for only 50 Myrs, as they require a timestep of $\sim$1 hour.  The initial orbits and masses assumed for 55 Cnc's planets are shown in Table 1.

In addition to the planets' and stars' mutual gravitational interactions, we also include external perturbations from the Galactic tide and passing field stars, since these forces affect the dynamics of 55 Cnc B for semimajor axes beyond a few thousand AU.  We employ the Galactic tidal field described in \citet{lev01}.  To model perturbations from passing stars, we use the impulse approximation \citep{rick76}.  These impluses are generated by randomly drawing from a stellar mass function \citep{reid02} and a mass-dependent velocity distribution \citep{garc01}.  

\section{Results and Discussion}

In Figure 1, we display the inclination evolution seen in one of our simulations.  This simulation is one of our 4-planet simulations, and 55 Cnc B is placed on an initial orbit with $a=1250$ AU, $e=0.93$, and $i=115^{\circ}$.  (Based projected separation, the most probable semimajor axis is $\sim$1340 AU \citep{fismar92}.) We see in this figure that the planets of 55 Cnc leave their initial orbital plane very rapidly.  In fact, within 30 Myrs the planets have reached a retrograde configuration relative to their original orbital plane.  If the parent star's spin axis is initially aligned with the planets' orbits, this means that the spin-orbit angle of the system quickly diverges from 0$^{\circ}$.  For this entire Gyr simulation, the planets continue to precess between prograde and retrograde configurations in a very regular manner.  Interestingly, the inset plot of this figure illustrates that all four planets maintain a very tight coplanar configuration throughout this evolution.  

The dynamical mechanism that drives this inclination evolution was first documented in \citet{inn97}.  In a system consisting of just one planet embedded within a binary star system, perturbations from the secondary star will cause the planetary orbit to undergo Kozai-like behavior, where the inclination and eccentricity of the planet oscillate exactly out of phase as the longitude of pericenter ($\omega$) circulates or librates \citep{koz62}.  The frequency of this oscillation is a function of the star's mass, semimajor axis, and eccentricity as well as the planet's semimajor axis.  However, when more planets are added to the system the behavior can change.  This is because the evolution of $\omega$ may no longer be dominated by the stellar companion's perturbations.  Instead, the mutual interactions between the planets can drive the evolution of the longitude of pericenter.  As long as the precession timescale of $\omega$ is much shorter than the Kozai timescale due to the binary's perturbations, the evolution of the planetary orbits will not resemble a Kozai resonance.  Indeed, integrations of the 55 Cnc planets in isolation indicate that $\omega$ of planet d circulates every 2.5 Myrs.  In contrast, integrations with only planet d and a binary companion ($a=1340$ AU, $e=0.95$) yield a much larger typical circulation period of 10$^{7-8}$ Myrs.  Thus, the Kozai mechanism will not operate, and the planets' eccentricities and mutual inclinations remain low.  

However, in addition to $\omega$, the longitude of ascending node ($\Omega$) also precesses in the reference frame of binary orbital plane \citep{inn97}.  Unless the binary star's orbital plane and the initial planetary orbital plane coincide, the precession of $\Omega$ in the binary's reference frame will translate to an inclination precession with respect to the initial planetary plane.  Although each planet would have a different $\Omega$ precession rate in isolation, the self-gravity of the system causes the planets to precess at a uniform rate \citep{tak08, bat11}.  Consequently, the system maintains a rigid, coplanar shape, even though its orbital plane can become greatly inclined to the star's spin axis (assuming this axis is perpendicular to the original planetary orbital plane).  

Figure 1 also shows that although our simulated system spends time in retrograde configurations, its inclination never exceeds a certain value, $i_{max}$ (in this case $i_{max}\simeq130^{\circ}$).  The reason for this is that the vertical component of the planetary orbital angular momentum, $L_z$, is always observed to be conserved in the reference frame of the binary orbital plane.  Another way of saying this is that as $\Omega$ precesses, the angle between the orbit normal of the planetary system and the orbit normal of the binary is fixed.  Thus, the inclination with respect to the original planetary orbital plane reaches a maximum after $\Omega$ has precessed 180$^{\circ}$ in the binary orbital frame.  In the reference frame of the original planetary orbital plane 

\begin{equation}
\cos{i_{max}}=\cos{2i_{bin}}
\end{equation}
where $i_{bin}$ is the inclination of the binary with respect to the planetary orbital plane.

In the above simulation, we do not include the innermost planet of the 55 Cnc system, planet e.  Because of its tiny orbital period, the computing costs to follow such an integration for 1 Gyr would be prohibitive.  However, this planet is only 0.1 AU from the next nearest planet (b), and it is subject to powerful perturbations from interplanetary gravitational interactions just as all the other planets are.  Consequently, it too should conform to the rigid body nature of the evolution displayed in Figure 1.  

Fortunately, verifying this does not actually require a 1-Gyr integration.  A few of our 4-planet simulations have inclination precession rates even faster than the system shown in Figure 1 (due to a lower pericenter and semimajor axis for the binary orbit).  These configurations with faster precession rates allow us to use shorter numerical simulations to verify that planet e follows this rigid body precession.  In Figure 2, we display one such case.  This figure actually displays inclinations from two different simulations: planet e's inclination in a 5-planet simulation and the inclinations of planets b, c, f, and d from a 4-planet simulation.  The same binary orbit is used in both simulations.  We see that the behavior of each system is very similar, with inclinations oscillating between 0$^{\circ}$ and $\sim$95$^{\circ}$.  In addition, the period of oscillation is nearly the same, although the 5-planet system precesses a little slower (due to the fact that planet e has a smaller ``natural" $\Omega$ precession rate and slows the mean rate down slightly).  Because planet e is so close to its parent star, we include general relativistic precession as well as a $J_2$ component to the parent star's potential ($J_2$ = 5 x 10$^{-7}$).  However, the  behavior of planet e does not change noticeably whether we include these effects or not.  Thus, we conclude that the inclination behavior seen in our 4-planet simulations is an excellent proxy for planet e's inclination as well.

Although we can use our 4-planet simulations to extrapolate the behavior of all 5 planets in 55 Cnc, this is still not an ideal computing situation.  Because of planet b's short orbital period, even our 4-planet simulations use a timestep of $\sim$1 day, and we would like to integrate our planetary systems for 10 Gyrs rather than 1 Gyr.  In addition, we would like to evolve many hundreds of systems, each with a different binary star orbit, since only the binary's projected separation is known.  

We now argue that 55 Cnc's inclination evolution can be modeled accurately using integrations that include just the outermost planet (d) accompanied by a fictitious inner planet (g), rather than the known configuration.  The reason for this is that planet d contains most of the total planetary mass of 55 Cnc.  Therefore, planet d dominates the system's self-gravity that maintains its rigid behavior.  The other planets should almost behave as test particles being dragged along with planet d.  The only role that the other planets play in this evolution is perturbing planet d's longitude of pericenter to prevent Kozai oscillations.  Consequently, the inclinations of any configuration of inner planets will evolve similarly as long as they are stable and perturb planet d's $\omega$ sufficiently.  Considering this, we choose to replace the inner 4 planets with a Saturn-mass planet at 3.5 AU (which we call planet g).  Such a planet is known to be stable \citep{ray08}, and its large semimajor axis enables us to increase our integration step from 1 day to 125 days.   

To demonstrate the accuracy of these new simulations, we compare our 48 4-planet simulations with simulations using planet g in place of b, c, and f.  For both our 4-planet simulations and our 2-planet simulations, we measure the maximum planetary inclination attained during the first Gyr.  These are plotted against each other in Figure 3a.  We see that there is a tight 1:1 correlation between the inclinations attained in both simulation sets.  Another parameter we can compare between the 2-planet and 4-planet systems is the precession rates of their planetary inclinations.  This is done in Figure 3b where we plot the precession periods (obtained from an FFT) from both simulation sets.  We see that in general the two precession periods are very near each other, although the tighter configuration of the 4-planet system does systemically yield a slightly lower precession rate.  Given that the inclination behaviors of these two simulation sets are nearly identical, we conclude that our 2-planet simulations are a suitable model for the inclination of the real 55 Cnc system.

In total we perform 450 2-planet simulations that are integrated for 10.2 Gyrs.  Each simulation includes a binary companion set on a different randomly generated orbit.  In a small fraction of our simulations (16\%), binary perturbations destabilize the planets, ejecting one or both.  Because this has not occurred in the real 55 Cnc system, we ignore these runs.  Using only the stable runs, we sample the inclination of planet d relative to its initial plane every Myr between 7 and 10.2 Gyrs.  The median values of those inclination samplings are plotted as a function of binary semimajor axis in the upper panel of Figure 4.  In addition, we use error bars to mark the boundaries of the upper 10\% and bottom 10\% of inclination measurements in each semimajor axis bin.  

If we assume that the parent star's spin was originally aligned with the planetary orbits, this plot shows the current spin-orbit angle of 55 Cnc.  We see that for binary semimajor axes below 4000 AU the most likely spin-orbit angle should be $\sim$65$^{\circ}$.  Beyond semimajor axes of $\sim$5000 AU, the precession timescale becomes much larger than the system's age, and the influence of the binary star wanes.  Hence, for the very largest binary separations, the planetary system is likely to be found only at low inclinations.  Lastly, we note that although our median inclination is always prograde, 29\% of the inclinations recorded in our simulations are retrograde for binaries with $a<5000$ AU.  

In the bottom panel of Figure 4, we show the cumulative probability distribution for the possible semimajor axis of 55 Cnc B.  To calculate this distribution, we assumed that the distribution of wide binary semimajor axes is uniform in log space \citep{pov07}.  Furthermore, we assumed that for a given semimajor axis, the distribution of all other orbital elements for wide binaries is isotropic.  Based on this orbital distribution, we measured the relative fraction of time that projected separations of 1065 AU occur.  We have seen in the upper panel that the binary companion has its strongest effects on the spin-orbit angle for $a<5000$ AU.  Our probability distribution in the bottom panel indicates that there is a $\sim$95\% probability that the semimajor axis of 55 Cnc B is within this range.  Thus, it is very likely that 55 Cnc B has significantly altered the spin-orbit angle of this system.

These simulations indicate that the current spin-orbit angle of 55 Cnc should be quite high and perhaps even retrograde.  This result is particularly exciting because 55 Cnc e was just recently discovered to be transiting \citep{winn11,dem11}.  During this planet's 95-minute transit, we estimate a 48 cm/s magnitude for the Rossiter-McLaughlin effect, assuming the star's maximum rotational velocity is damped by the planet-to-stellar radius ratio squared \citep{vonb11, gaudwinn07}.  This is slightly above current detection limits \citep{may11}, and it should be possible to measure the spin-orbit angle.  Because the measured angle is only projected, we also display the range of the projected angle ($\lambda$) in the upper panel Figure 4,  assuming a random orientation on the sky.  Although the median values of $\lambda$ are lower than the true inclinations, they are still well above the 30$^\circ$ ``misaligned threshold" specified in \citet{tri10} for all binary semimajor axes besides $a=8000$ AU.

\section{Conclusions}

We show that it is very likely that perturbations from 55 Cnc B have significantly altered the spin-orbit angle of the planetary system of 55 Cnc A, even though the planetary system's self-gravity has preserved the planets' initial coplanarity.  Hence, even very distant binary companions such as 55 Cnc B can substantially alter planetary architectures.  Assuming wide binary orbits reflect an isotropic distribution, we demonstrate in Figure 4 that the most likely value of 55 Cnc's projected spin-orbit angle is $\sim$45--55$^{\circ}$.  Furthermore, there is a significant (29\%) chance of a retrograde configuration for 55 Cnc.  Because 55 Cnc is a closely packed, coplanar system, it would be unique among known highly inclined planetary systems.  This feature could distinguish retrograde planets produced from rigid body precession vs. those generated from Kozai interactions and/or scattering events.  Since planet e of this system transits its parent star, it may be possible to soon verify our findings using the Rossiter-McLauglin effect.

\section{Acknowledgements}
We thank John Chambers for valuable discussions concerning the MERCURY package.  This work was funded by a CITA National Fellowship and Canada's NSERC.  SNR thanks the CNRS's PNP program and the NASA Astrobiology Institute's Virtual Planetary Laboratory team.  The bulk of our computing was performed on SciNet's General Purpose Cluster provided at the University of Toronto.
\clearpage

\begin{table}[htbp]
\centering
\begin{tabular}{c c c c c}
\hline
Planet & $M$ & $a$ & $e$ & $i$ \\
 & (M$_{Jup}$) & (AU) & & ($^{\circ}$)\\
\hline
e & 0.027 & 0.016 & 0.001\footnotemark & 0--1\footnotemark\\
b & 0.83 & 0.11 & 0.01 & 0--1\\
c & 0.17 & 0.24 & 0.005 & 0--1\\
f & 0.16 & 0.78 & 0.30 & 0--1\\
{\it g} & {\it 0.29} & {\it 3.5} & {\it 0.001} & {\it 0--1}\\
d & 3.82 & 5.74 & 0.014 & 0--1\\
\hline
\end{tabular}
\caption{List of planet masses, semimajor axes, eccentricities, and inclinations used in our simulations' initial conditions \citep{dawfab10}.  All other unlisted orbital elements were randomly drawn from a uniform distribution from 0 to 2$\pi$.  Note that planet g is a stable fictitious planet added to some of our runs.  }
\end{table}

\setcounter{footnote}{1}
\footnotetext{Although \citet{dawfab10} list the eccentricity of planet e as 0.06, their analysis is also consistent with zero eccentricity, and it is likely such a closely orbiting planet would be tidally circularized rapidly.  We have run a subset of alternative simulations with an eccentric planet e, and our results are independent of this orbital parameter.}
\setcounter{footnote}{2}
\footnotetext{To avoid 2-D integrations, we give each planet a slightly different random initial inclination between 0 and 1 degree.}

\begin{figure}[tbp]
\centering
\includegraphics[scale=.6]{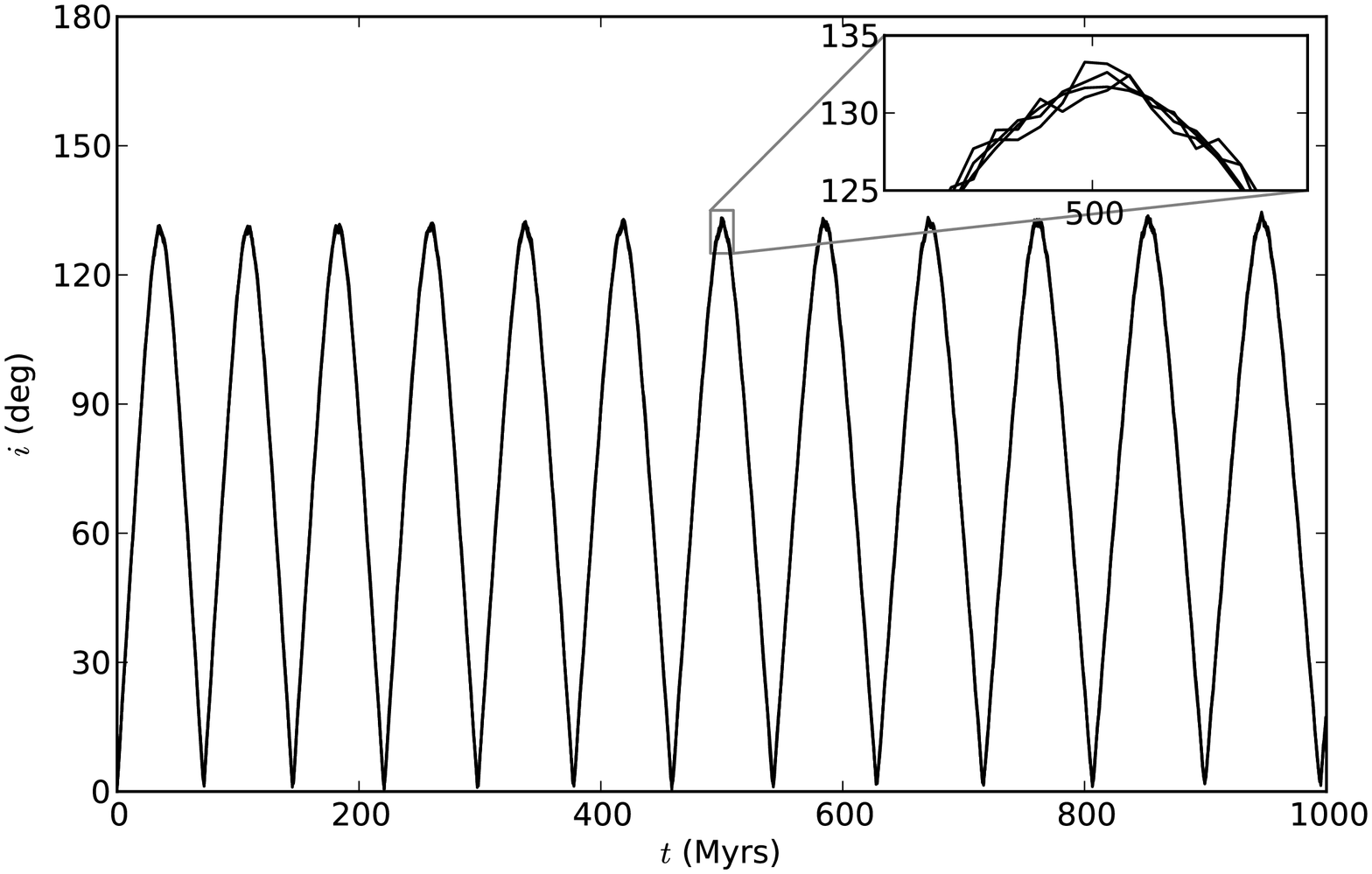}
\caption{Evolution of the inclination of the outer four planets of the 55 Cancri system vs. time.  Inclination is measured relative to the initial planetary orbital plane.  The inset plot resolves the evolution each planet's inclination.}\label{fig:samp}
\end{figure}
\clearpage

\begin{figure}[tbp]
\centering
\includegraphics[scale=.6]{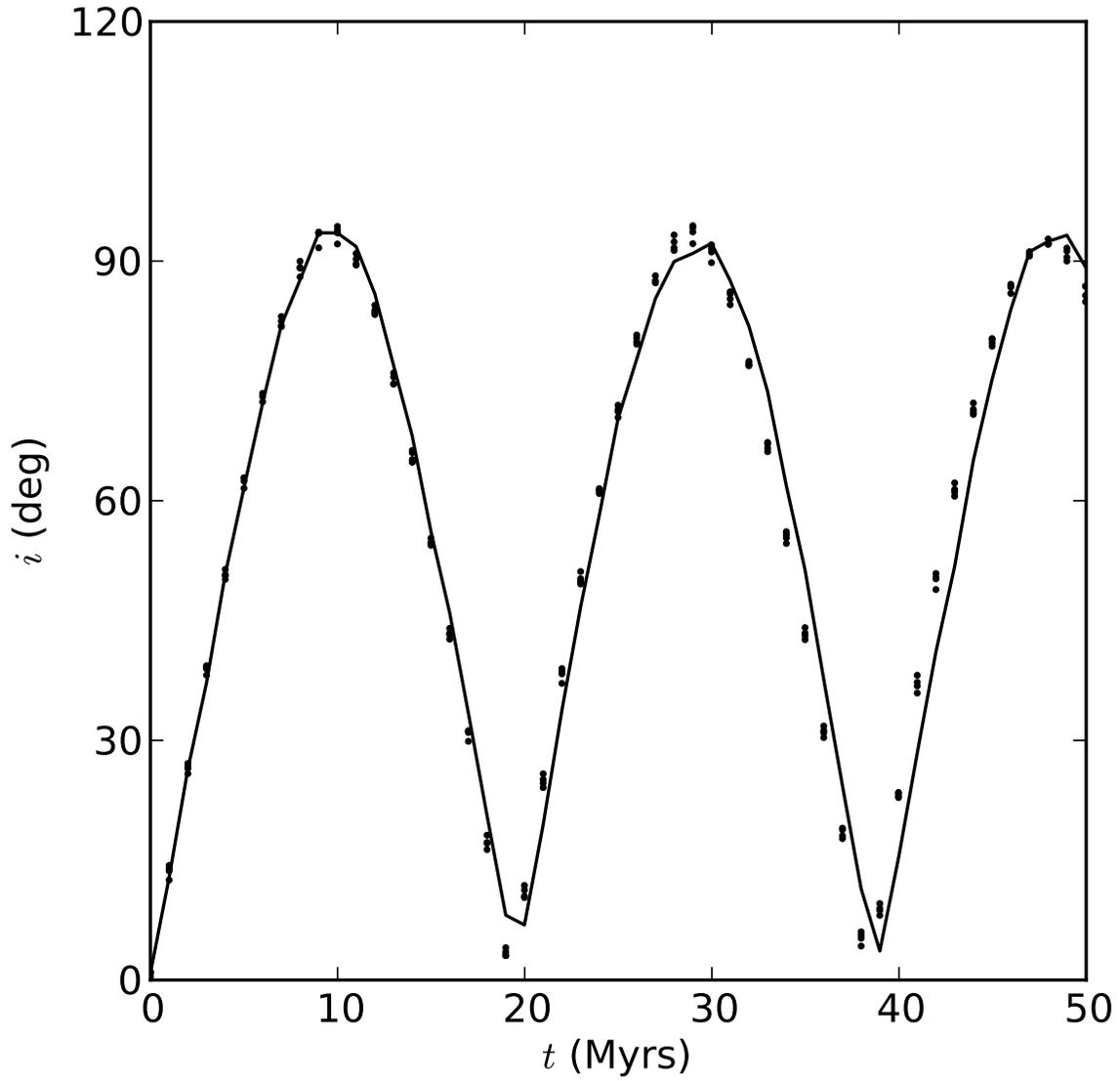}
\caption{Inclination evolution of the 55 Cancri planets in two different simulations using the same binary orbit.  The solid line shows the evolution of planet e in a simulation including all 5 planets.  The data points plot the inclinations of the outer four planets in a simulation including just those planets.  Inclinations are measured relative to the initial planetary orbital plane.}\label{fig:addpl}
\end{figure}
\clearpage

\begin{figure}[tbp]
\centering
\includegraphics[scale=.8]{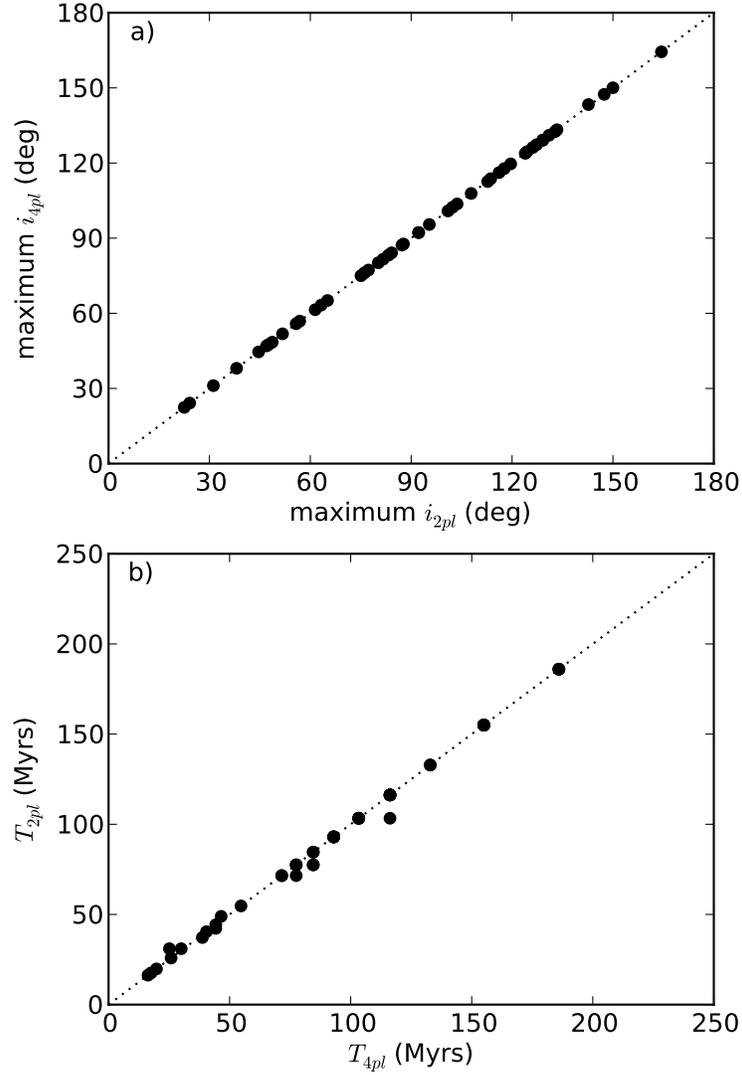}
\caption{{\it \bf a:} Maximum planetary inclination measured in 4-planet simulations vs. the maximum planetary inclination measured in 2-planet simulations.  Inclinations are measured relative to the initial planetary orbital plane.  {\it \bf b:} The inclination precession period measured in 2-planet simulations vs. the precession period measured in 4-planet simulations. Data for both plots only come from the first Gyr of each simulation.  }\label{fig:percomp}
\end{figure}
\clearpage

\begin{figure}[tbp]
\centering
\includegraphics[scale=.8]{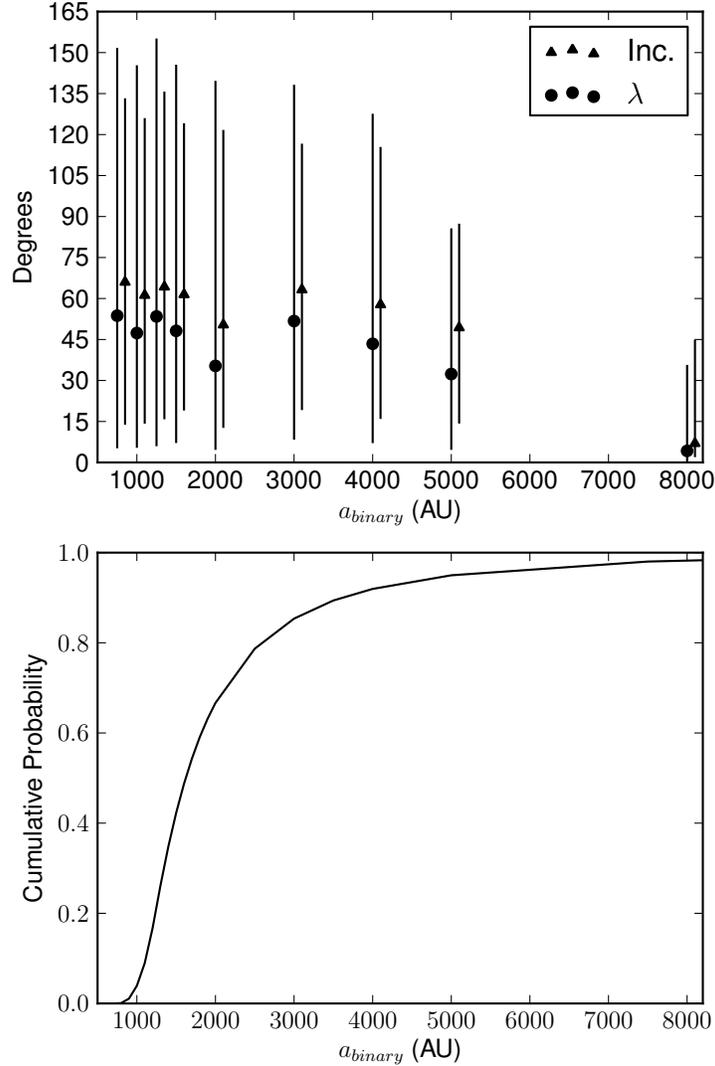}
\caption{{\it \bf Upper panel:} The predicted range of the spin-orbit angle of the 55 Cnc planetary system as a function of the system's binary semimajor axis.  The circular data points mark the median value of the true spin-orbit angle, and the triangular data points mark the median value of the projected angle ($\lambda$) that could be measured via the R-M effect.  The error bars mark the range between the lower 10\% and upper 90\% of all possible angle values seen in our simulations.  The ranges of the projected angle and true angle are offset slightly at each binary semimajor axis to avoid confusion. {\it \bf Bottom panel:} Cumulative probability distribution of the possible semimajor axis of 55 Cnc B if its projected separation distance is 1065 AU.  This distribution assumes that wide binary semimajor axes are uniformly distributed in $\log{a}$-space.  All other binary orbital elements are assumed to reflect an isotropic distribution.  }\label{fig:predict}
\end{figure}
\clearpage


\end{document}